\documentclass[11pt,a4paper,english,superscriptaddress,aps]{revtex4}
\usepackage{amsmath,amssymb}
\makeatletter
\usepackage{babel}

\usepackage{hyperref}

\renewcommand{\a}{\alpha}
\renewcommand{\b}{\beta}

\newcommand{\pa}{\partial}

\begin{document}

\title{The three-dimensional non-anticommutative superspace}

\author{A. F. Ferrari}
\author{M. Gomes}
\affiliation{Instituto de F\'\i sica, Universidade de S\~ao Paulo\\
Caixa Postal 66318, 05315-970, S\~ao Paulo, SP, Brazil}
\email{alysson,mgomes,ajsilva@fma.if.usp.br}
\author{J. R. Nascimento} 
\author{A. Yu. Petrov}
\affiliation{Departamento de F\'{\i}sica, Universidade Federal da Para\'{\i}ba\\
 Caixa Postal 5008, 58051-970, Jo\~ao Pessoa, Para\'{\i}ba, Brazil}
\email{jroberto,petrov@fisica.ufpb.br}
\author{A. J. da Silva}
\affiliation{Instituto de F\'\i sica, Universidade de S\~ao Paulo\\
Caixa Postal 66318, 05315-970, S\~ao Paulo, SP, Brazil}
\email{mgomes,ajsilva@fma.if.usp.br}

\begin{abstract}
We propose two alternative formulations for a three-dimensional non-anticommutative superspace in
which some of the fermionic coordinates obey  Clifford
anticommutation relations. For this superspace, we construct the
supersymmetry generators satisfying  standard anticommutation
relations and the  corresponding supercovariant derivatives. We
formulate a scalar superfield theory in such a superspace and
calculate its propagator. We also suggest a prescription for the introduction of interactions  in such theories. 
\end{abstract}

\maketitle

The concept of noncommutative coordinates has deep motivations originated
from  fundamental properties of the space-time and string
theory \cite{SW}. In this context a recent and very important development was the introduction of a 
non-anticommutative superspace \cite{Klemm,Seiberg} in which the noncommutativity affects not only the
usual space-time coordinates, like in \cite{SW}, but also the
fermionic coordinates. Such an idea was shown to have
some foundations in string theory \cite{SB}.  

In the four-dimensional case, the superspace in which the fermionic
coordinates are non-anticommutative   possesses a
very specific form of supersymmetry, called $N=\frac12$
supersymmetry \cite{Seiberg,FL}. This fact generated a great deal  of interest in  field
theories formulated in the four-dimensional non-anticommutative superspace. 
Among the results obtained in the ensuing investigations, we would like to mention
the proof of the renormalizability of the
non-anticommutative Wess-Zumino model \cite{Roma} and the 
development of the background field method for the
super-Yang-Mills theory \cite{Penati}.
However, until this time the  non-anticommutative superspace formulation for the
three-dimensional space-time has not been discussed, despite
the fact that  supersymmetry in three dimensions is
very simple (see its detailed description in \cite{SGRS};
interesting examples of noncommutative three-dimensional superfield
theories  can be found in \cite{sigma,cpn,ours,chern}). 
The main reason for such  situation seems to be the following: the relative
simplicity of the four-dimensional noncommutative superspace
is based on the fact that one can keep untouched half of the supersymmetry generators, which obeys standard
anticommutation relations. This possibility arises because the Lorentz group has two independent fundamental spinor representations (the dotted and the undotted ones), and correspondingly there are two sets of supersymmetry generators. In this case, we have the choice of imposing
nontrivial anticommutation relations between the fermionic
coordinates affecting the algebra of only one of the sets (the dotted ones, let us say), preserving the algebra of the other set (namely, the undotted ones)~\cite{foot2}. In the three-dimensional space-time,
however, the lowest dimensional spinor representation of the Lorentz group is real, so there is only one set of supersymmetry generators. Hence, imposing nontrivial anticommutation relations between the fermionic
coordinates could break completely the supersymmetry algebra. Although this possibility does not necessarily lead to inconsistencies, in this work we require that at least some explicit supersymmetry survive. Hence, we have to develop alternative ways to introduce non-anticommutativity in the three-dimensional superspace.

Starting with $N=1$ supersymmetry we may define modified supersymmetry generators satisfying the standard anticommutation relations but constructing the field theories would be problematic. We will return to this point later. For this reason, in this paper we begin with the $N=2$ supersymmetry. So
the starting point of the three-dimensional non-anticommutative superspace
is the following anticommutation relation for the fermionic
coordinates $\theta^{i\alpha}$:
\begin{eqnarray}
\label{basic}
\{\theta^{i\alpha},\theta^{j\beta} \}=\Sigma^{ij\alpha\beta}.
\end{eqnarray}

\noindent
Here $\alpha,\beta=1,2$ are Lorentz spinor indices, $i,j=1,2$ are labels for the fermionic coordinates corresponding to each supercharge, and $\Sigma^{ij\alpha\beta}$ is a constant matrix symmetric under the exchange of $i \alpha$ and $j \beta$. Let us construct the supersymmetry
generators for this case. For simplicity and for the sake of
clarification of the impact of this relation, we assume that all
non(anti)commutativity is concentrated in the fermionic coordinate
sector, the bosonic space-time coordinates being assumed to commute.

In the case of non-anticommutative superspace, the usual supersymmetry generators (here
$\partial^i_{\alpha}=\frac{\partial\phantom a}{\partial\theta^{i\alpha}}$, we use the notations of \cite{SGRS})
\begin{eqnarray}
\label{usualQ}
Q^i_{\alpha}=i\partial^i_{\alpha}+\theta^{i\beta}\partial_{\beta\alpha},
\end{eqnarray}

\noindent
have the following anticommutation relation (we assume that
$\{\partial^i_{\a},\partial^j_{\b}\}=0$):
\begin{eqnarray}
\{Q^i_{\alpha},Q^j_{\beta}\}=
2P_{\alpha\beta}\delta^{ij}-\Sigma^{ij\gamma\delta}P_{\gamma\alpha}P_{\delta\beta},
\end{eqnarray}

\noindent
where $P_{\alpha\beta}=i\partial_{\beta\alpha}$ are the momentum generators. This
relation differs in an essential way from the general assumptions of
standard supersymmetric theories since it is nonlinear, implying that the
anticommutator of two transformations generated by the $Q$'s is not a translation but a more complicated transformation. 

At first sight one could try to overcome this difficulty by finding
new supersymmetry generators,
$\tilde{Q}^i_{\alpha}$,
which would satisfy the standard anticommutation relation
\begin{eqnarray}
\{\tilde{Q}^i_{\alpha},\tilde{Q}^j_{\beta}\}=2\delta^{ij}P_{\alpha\beta}.
\end{eqnarray}
It is easily verified  that this  relation is satisfied  if
\begin{eqnarray}
\label{newgen}
\tilde{Q}^i_{\alpha}=Q^i_{\alpha}+\frac{i}{2}\Sigma^{ij\beta\gamma}\partial^j_{\beta}P_{\gamma\alpha}.
\end{eqnarray}

\noindent
However,  in the next step, to fix a (new) supercovariant
derivative $\tilde{D}^i_{\alpha}$ this approach meets a stumbling block. In fact, the condition of supersymmetric
covariance demands
\begin{eqnarray}
\{\tilde{D}^i_{\alpha},\tilde{Q}^j_{\beta}\}=0.
\end{eqnarray}

\noindent
Taking into account the basic relation
(\ref{basic}) and  the expression (\ref{newgen}) for the 
supersymmetry generator, a natural candidate would be
\begin{eqnarray}
\label{newgen1}
\tilde{D}^i_{\alpha}=D^i_{\alpha}+\frac{1}{2}\Sigma^{ij\beta\gamma}
\partial^j_{\beta}P_{\gamma\alpha},
\end{eqnarray}

\noindent
where $D^i_{\a}=\partial^i_{\alpha}+i\theta^{i\beta}\partial_{\beta\alpha}$
is the standard supercovariant derivative. 
Nevertheless, these operators involve second derivatives and do not satisfy Leibnitz's rule, so one cannot use integration by parts, which turns cumbersome the D-algebra transformations~\cite{foot1}.
Had we insisted in the formulation of a non-anticommutative superspace beginning with $N=1$ supersymmetry, we would arrive at the same difficulty. This is the problem mentioned earlier.

However, since we started with $N=2$, there are alternative ways to introduce the non-anticommutativity. We will describe two of them. Our first proposal is similar to the conception of the $N=(1,1)$ superspace \cite{Ivanov} in four dimensional superspace. We break half of supersymmetries, that
is, those corresponding to the generators $Q^2_{\a}$: we choose
 $\Sigma^{11\a\b}=\Sigma^{21\a\b}=\Sigma^{12\a\b}=0$, so that the only nontrivial
anticommutation relation for the fermionic coordinates is
\begin{eqnarray}
\label{basic2}
\{\theta^{2\alpha},\theta^{2\beta} \}=\Sigma^{\alpha\beta},
\end{eqnarray}

\noindent
hence only the unbroken generators $Q^1_{\alpha}$ satisfy the standard
anticommutation relation
\begin{eqnarray}
\{{Q}^1_{\alpha},{Q}^1_{\beta}\}=2i\pa_{\alpha\beta}.
\end{eqnarray}

The supercovariant derivatives defined to anticommute with these
generators are \cite{foot}
\begin{eqnarray}
D^i_{\a}=\pa^i_{\a}+i\theta^{i\b}\pa_{\b\a},
\end{eqnarray}
and one can verify that
\begin{eqnarray}
\label{c1}
\{D^1_{\a},D^1_{\b}\}=2i\pa_{\a\b}, \quad\,\{D^1_{\a},D^2_{\b}\}=0,
\quad\,
\{D^2_{\a},D^2_{\b}\}=2i\pa_{\a\b}-\Sigma^{\gamma\delta}\pa_{\gamma\a}
\pa_{\delta\b}.
\end{eqnarray}

The most natural definition of the Moyal product compatible with
the condition of supersymmetric covariance is
\begin{eqnarray}
\label{moyal0} \Phi_1(z)*\Phi_2(z)=\exp\left[-\frac{1}{2}\Sigma^{\a\beta}D^2_{\a}(z_1)
D^2_{\b}(z_2)\right]\Phi_1(z_1)\Phi_2(z_2)|_{z_1=z_2=z}.
\end{eqnarray}

\noindent
This form is similar to the Moyal product introduced in \cite{Seiberg}. 
We note the  presence of  supercovariant
derivatives in the exponent instead of ordinary derivatives $\partial_{\alpha}$.
This is so because ordinary spinor derivatives do not anticommute with the
supersymmetry generators ${Q}^1_{\alpha}$ and their use in
(\ref{moyal0}) would lead to more complicated transformation rules.
We note also that the sign of  the exponent is fixed by the
condition that the (Moyal) anticommutator
$\{\theta^{2\,\alpha},\theta^{2\,\beta}\}_*$ must reproduce the relation
(\ref{basic2}). Expanding the exponential in (\ref{moyal0}) we obtain,
\begin{eqnarray}
\label{defmoyal} 
&&\exp\left[-\frac{1}{2}\Sigma^{\alpha\beta}D^2_{\a}(z_1)D^2_{\beta}(z_2)\right]
\Phi_1(z_1)*\Phi_2(z_2)|_{z_1=z_2=z} \nonumber\\
&&\equiv\left[1-\frac{1}{2}\Sigma^{\alpha\beta}D^2_{\alpha}(z_1)D^2_{\b}(z_2)+\frac{1}{8}\Sigma^{\alpha\beta}
D^2_{\a}(z_1)D^2_{\beta}(z_2)\Sigma^{\gamma\delta}
D^2_{\gamma}(z_1)D^2_{\delta}(z_2)+\ldots\right]\Phi_1(z_1)\Phi_2(z_2)|_{z_1=z_2=z}\nonumber\\
&&=
\Phi_1(z)\Phi_2(z)-\frac{1}{2}\Sigma^{\a\beta}D^2_{\alpha}\Phi_1(z) D^2_{\beta}\Phi_2(z)-
\frac{1}{8}\Sigma^{\a\beta}\Sigma^{\gamma\delta}D^2_{\alpha}D^2_{\gamma}\Phi_1(z)
D^2_{\beta}D^2_{\delta}\Phi_2(z)+\ldots.
\end{eqnarray}

\noindent
A distinct feature of (\ref{defmoyal}) is that, unlike  the
four-dimensional case, it involves all orders in $D^2_{\alpha}$,
since there is no power of the supercovariant derivative in
three dimensions which identically vanishes. For the Moyal product of three
or more superfields this definition can be straightforwardly
generalized. 

We are now in a position to define the propagators for field theories in the non-anticommutative
superspace. We consider the simplest example, i.e., the scalar superfield theory.
The  most natural form of its free action is
\begin{eqnarray}
\label{quad1}
S=-\frac{1}{4}\int d^7 z\,\Phi*\Delta\Phi,
\end{eqnarray}
where $d^7z=d^3x d^2\theta^1 d^2\theta^2$ and $\Delta$ is some
operator which does not depend on $\theta^2$ so that it is not affected by the Moyal product.

The reader should not be misled by the apparent simplicity of this action.
Indeed, in contrast with the four-dimensional case, the above expression nontrivially involves the noncommutativity matrix $\Sigma^{\a \b}$! 
In other words, the superspace integral of the Moyal product of two fields essentially differs from the usual case. Really, it follows from (\ref{defmoyal}) that
\begin{eqnarray}
S&=&\frac{1}{2}\int d^7z
\Big(\Phi(z)\Delta\Phi(z)-\frac{1}{2}\Sigma^{\a\beta}D_{\alpha}\Phi(z)
D_{\beta}\Delta\Phi(z)-\nonumber\\&-&
\frac{1}{8}\Sigma^{\a\beta}\Sigma^{\gamma\delta}D_{\alpha}D_{\gamma}\Phi(z)
D_{\beta}D_{\delta}\Delta\Phi(z)+\ldots\Big).
\end{eqnarray}

\noindent
By moving all derivatives to the field affected by
$\Delta$ by means of integration by parts, we get
\begin{eqnarray}
S&=&\frac{1}{2}\int d^7z
\Big(\Phi(z)\Delta\Phi(z)+\frac{1}{2}\Sigma^{\a\beta}\Phi(z)
D^2_{\alpha}D^2_{\beta}\Delta\Phi(z)+\nonumber\\&+&
\frac{1}{8}\Sigma^{\a\beta}\Sigma^{\gamma\delta}\Phi(z)
D^2_{\gamma}D^2_{\alpha}D^2_{\beta}D^2_{\delta}\Delta\Phi(z)+\ldots\Big).
\end{eqnarray}

\noindent
In general, the $n$th order in $\Sigma$ term has the form
\begin{eqnarray}
\label{term} 
A_n=\frac{1}{2^{n+1} \, n!}\int d^7 z \,
\Phi(z)\Sigma^{\a_n\b_n}\dots\Sigma^{\a_2\b_2}\Sigma^{\a_1\b_1}
D^2_{\a_n}D^2_{\a_{n-1}}\ldots D^2_{\a_1}D^2_{\b_1}\ldots
D^2_{\b_n}\Delta\Phi(z).
\end{eqnarray}
The crucial step here is a very specific grouping of derivatives
which can be efficiently simplified. Indeed, as
$D^2_{\a_1}$ and $D^2_{\b_1}$ in this expression are adjacent and are contracted with the symmetric matrix
$\Sigma^{\a_1\b_1}$, we can use (\ref{c1}) to replace
$D^2_{\a_1}D^2_{\b_1}$ by
$\frac{1}{2}\{D^2_{\a 1},D^2_{\b 1}\}=
P_{\a_1\b_1}+\frac{1}{2}\Sigma^{\gamma\delta}P_{\gamma\a_1}P_{\delta\b_1}$.
We call the combination $\Sigma^{\a_1\b_1}D^2_{\a_1}D^2_{\b_1}$ as
\begin{eqnarray}
\label{r} R(P)\equiv\Sigma^{\a\beta}(P_{\a\beta}+\frac{1}{2}
\Sigma^{\gamma\delta}P_{\gamma\alpha}P_{\delta\beta})=
2\Sigma\cdot P +2(\Sigma\cdot P)^2-\Sigma^2P^2\,,
\end{eqnarray}

\noindent
where we have introduced the vector
$\Sigma^n=\frac{1}{2}(\gamma^n)^{\a\beta}\Sigma_{\a\beta}$, and $\Sigma\cdot
P=\Sigma^m P_m$. Then, since $R(P)$ commutes with supercovariant derivatives, we can factor it out 
from (\ref{term}) and consider the next pair of
derivatives, contracted as $\Sigma^{\a_2\b_2}D_{\a_2}D_{\b_2}$. As
a result we obtain another factor of $R(P)$. After repeating this procedure
$n$ times, we eventually arrive at
\begin{eqnarray}
A_n=\frac{1}{2\,n!} \int
d^7 z\, \Phi(z) \left[\frac{R(P)}{2}\right]^n \Delta\Phi(z).
\end{eqnarray}

\noindent
Using this result, the quadratic action (\ref{quad1}) takes the form,
\begin{eqnarray}
S=\frac{1}{2}\int d^7 z\, \Phi(z)\,e^{R(P)/2} \, \Delta\Phi(z).
\end{eqnarray}
Notice that, differently from the space-time noncommutativity, the above quadratic action is deeply affected by the Moyal product. 
The corresponding momentum space propagator reads
\begin{eqnarray}
\label{prop}
 <\Phi(-p,\theta_1)\Phi(p,\theta_2)>\,=\,ie^{-R(P)/2}\Delta^{-1}(p)\delta^4_{12}.
\end{eqnarray}

\noindent
One possibility to avoid the blow up of the propagator with the growth of the momentum is to choose the $\Sigma^m$ vector to be light-like, $\Sigma^2=0$, which gives $<\Phi(-p,\theta_1)\Phi(p,\theta_2)>\,=\,ie^{-\Sigma\cdot p -(\Sigma\cdot p)^2} \Delta^{-1}(p)\delta^4_{12}$, thus implying in a fast (exponential) decrease with the momentum $p$ along certain directions. A more interesting possibility is to choose $\Sigma^m$ to be time-like, $\Sigma^m=(\Sigma_0,0,0)$, which gives $<\Phi(-p,\theta_1)\Phi(p,\theta_2)>\,=\,ie^{\Sigma_0 p_0 -\Sigma_0^2(p_0^2+\vec{p}^2)} \Delta^{-1}(p)\delta^4_{12}$ with exponential decrease along all directions; in the case of space-like $\Sigma^m$ the theory is unstable.
In all cases the propagator reflects the fact that the quadratic action is highly nonlocal. 
 
The interaction vertices are defined by a direct generalization of (\ref{moyal0}) for arbitrary
number of fields,
\begin{eqnarray}
&&\Phi_1(z)*\Phi_2(z)*\ldots*\Phi_n(z)=\nonumber\\
&&=\exp\left[-\frac{1}{2}\sum_{i\leq j\leq n}
\Sigma^{\a\beta}D^2_{\a}(z_i)D^2_{\b}(z_j)\right]\Phi_1(z_1)\Phi_2(z_2)\ldots\Phi_n(z_n)|_{z_1=z_2=\ldots
z_n=z}.
\end{eqnarray}

\noindent
The perturbative series should be ordered by powers of the $\Sigma$ that appears in the interaction terms (the propagator must not be expanded). This procedure, in the case of the time-like $\Sigma^m$ guarantees the order by order finiteness of this expansion.

A natural question which could arise in connection with our construction concerns the role played by the supersymmetry generated by the $Q^1$ supercharge. To clarify further this point let us consider the following example. The $N=2$ spinor superfield $\Psi_{\alpha}(z)$ can be expanded in terms of the $N=1$ superfields as:
\begin{eqnarray}
\label{expsi}
\Psi_{\alpha}(z)=\lambda_{\alpha}(x,\theta^1)+\theta^2_{\alpha}\Phi(x,\theta^1)+\theta^{2\beta}A_{\alpha\beta}(x,\theta^1)+
(\theta^2)^2 F_{\alpha}(x,\theta^1),
\end{eqnarray}
where $\Phi(x,\theta^1)$, $\lambda_{\alpha}(x,\theta^1),A_{\alpha\beta}(x,\theta^1),F_{\alpha}(x,\theta^1)$ 
are $N=1$ superfield components of the $N=2$ supermultiplet, and $A_{\alpha\beta}(x,\theta^1)$ is a symmetric bispinor. 
If we introduce the action for such a superfield as being
\begin{eqnarray}
S=-\frac{1}{4}\int d^7 z\,\Psi^{\alpha}(z)*(D^1)^2\Psi_{\alpha}(z),
\end{eqnarray}
which is a special case of the Eq. (\ref{quad1}), we can apply the arguments given above to show that this action can be rewritten as
\begin{eqnarray}
-\frac{1}{4}\int d^7 z\, \Psi^{\alpha}(z)\,e^{R(P)/2} \, (D^1)^2\Psi_{\alpha}(z).
\end{eqnarray}
Substituting here the expansion (\ref{expsi}), we arrive at the following action:
\begin{eqnarray}
&-&\frac{1}{2}\int d^7 z\, (\theta^2)^2\,\Big[\Phi(x,\theta^1) e^{R(P)/2}\, (D^1)^2\Phi(x,\theta^1)+
2\lambda^{\alpha}(x,\theta^1)e^{R(P)/2}(D^1)^2F_{\alpha}(x,\theta^1)+\nonumber\\&+&A^{\alpha\beta}(x,\theta^1)e^{R(P)/2}(D^1)^2 A_{\alpha\beta}(x,\theta^1)\Big],
\end{eqnarray}
so the integral over $d^2\theta^2$ is factorized out. Hence, after integration over $\theta^2$ the action takes the form
\begin{eqnarray}
&&\frac{1}{2}\int d^5 z\,\Big[\Phi(x,\theta^1) e^{R(P)/2}\, (D^1)^2\Phi(x,\theta^1)+
2\lambda^{\alpha}(x,\theta^1)e^{R(P)/2}(D^1)^2F_{\alpha}(x,\theta^1)+\nonumber\\&+&
A^{\alpha\beta}(x,\theta^1)e^{R(P)/2}(D^1)^2 A_{\alpha\beta}(x,\theta^1)\Big] ,
\end{eqnarray}
and, as a consequence, we have generated the exponential factor for each $N=1$ superfield component of the $N=2$ supermultiplet. We conclude that, despite the second supersymmetry is being explicitly broken, its nontrivial impact persists even after integration over the corresponding spinor coordinates. In other words, the $N=1$ theory still ``remembers'' the second supersymmetry.

Our second proposal to introduce non-anticommutativity for fermionic coordinates is the following: we first introduce the $N=2$ supersymmetry with the nontrivially mixed generators:
\begin{eqnarray}
\label{usualQ0}
Q^i_{\alpha}=i\partial^i_{\alpha}+(\sigma_1)^{ij}\theta^{j\beta}\partial_{\beta\alpha}.
\end{eqnarray}
To mix two supersymmetries in a nontrivial way, we have introduced the symmetric object $(\sigma_1)^{ij}$, where 
$\sigma_1=
\left(\begin{array}{cc}
0&1\\
1&0
\end{array}\right)$ is a Pauli matrix. The explicit form of the new generators is
\begin{eqnarray}
\label{usualQ1}
Q^1_{\alpha}=i\partial^1_{\alpha}+\theta^{2\beta}\partial_{\beta\alpha},\quad\,
Q^2_{\alpha}=i\partial^2_{\alpha}+\theta^{1\beta}\partial_{\beta\alpha}.
\end{eqnarray}
In the commutative case, their anticommutation relation is
\begin{eqnarray}
\{Q^i_{\alpha},Q^j_{\beta}\}=2i(\sigma_1)^{ij}\partial_{\alpha\beta}.
\end{eqnarray}
The covariant derivatives $D^i_{\alpha}$ which anticommute with these generators, $\{D^i_{\alpha},Q^j_{\beta}\}=0$, are
\begin{eqnarray}
\label{usualD}
D^i_{\alpha}=\partial^i_{\alpha}+i(\sigma_1)^{ij}\theta^{j\beta}\partial_{\beta\alpha}.
\end{eqnarray}
Now let us partially break this supersymmetry by imposing the following nontrivial anticommutator for spinor coordinates:
\begin{eqnarray}
\label{basic20}
\{\theta^2_{\alpha},\theta^2_{\beta}\}=\Sigma_{\alpha\beta}.
\end{eqnarray}
As a result, $\{Q^1_{\alpha},Q^1_{\beta}\}=\Sigma^{\gamma\delta}\partial_{\gamma\alpha}\partial_{\delta\beta}$ is deformed while other anticommutators of the generators will remain unchanged. It is easy to see that the $D^i_{\alpha}$ derivatives, both for $i=1,2$ still anticommute with the unbroken generators $Q^2_{\alpha}$. We note that in this case we cannot factorize two supersymmetries.

The derivatives $D^i_{\alpha}$ satisfy the following anticommutation relations:
\begin{eqnarray}
\label{antic}
\{D^2_{\alpha},D^2_{\beta}\}=0,\quad\,\{D^1_{\alpha},D^2_{\beta}\}=-2i\partial_{\alpha\beta},\quad\,
\{D^1_{\alpha},D^1_{\beta}\}=-\Sigma^{\gamma\delta}\partial_{\alpha\gamma}\partial_{\beta\delta}.
\end{eqnarray}
In this case, the most natural definition of the Moyal product compatible with
the condition of supersymmetric covariance is
\begin{eqnarray}
\label{moyal01} \Phi_1(z)*\Phi_2(z)=\exp\left[-\frac{1}{2}\Sigma^{\alpha\beta}D^2_{\alpha}(z_1)
D^2_{\beta}(z_2)\right]\Phi_1(z_1)\Phi_2(z_2)|_{z_1=z_2=z}.
\end{eqnarray}
This form is similar to the Moyal product introduced in \cite{Seiberg}. 
For exact the same reason as in~(\ref{moyal0}), only  supercovariant
derivatives are allowed in the exponent, and the sign of the exponent is fixed for Eq~(\ref{moyal01}) to be compatible with~(\ref{basic2}).
Similarly to~\cite{Seiberg}, in this case the Moyal product is a finite power series because third and higher degrees of $D^2_{\beta}$ are equal to zero due to (\ref{antic}). Also, we can verify that the quadratic action of any non-anticommutative theory will coincide with its anticommutative analog, that is, $\int d^7 z \Phi_1*\Phi_2=\int d^7 z\theta\Phi_1\Phi_2$. Hence, in this case, the propagators will not be changed after the non-anticommutative deformation, only vertices will be affected by the non-anticommutativity, which implies in the arising of a finite number of extra terms. For example, the non-anticommutative analog of the $\Phi^3$ vertex will take the form
\begin{eqnarray}
\int d^7 z \Phi*\Phi*\Phi=\int d^7 z \left\{\Phi^3- (\det \Sigma) \Phi \left[(D^1)^2\Phi\right]^2\right\}.
\end{eqnarray}
Other field theories in this non-anticommutative superspace can be analogously defined.

Let us now summarize our results. We proposed two alternative formulations for the three-dimensional
non-anticommutative superspace. This required the construction of a new representation of the 
supersymmetry algebra, which was realized by starting with a $N=2$ algebra, which was explicitly broken down to $N=1$ in two different ways. By using the supercovariant derivatives, we constructed a Moyal product compatible with the remaining supersymmetry. Unlike the four-dimensional case, in the first formulation,
the Moyal product is an infinite power series in the
noncommutativity matrix $\Sigma^{\a \b}$. Furthermore, the quadratic action is affected by the
noncommutativity. As a result, the propagator has a very unusual form characterized, for time-like $\Sigma^m$, by its exponential decay with the momentum, yielding very good convergence properties for the Feynman integrals. 
In particular, it implies that the
problem of UV/IR mixing \cite{Minw}, crucial in the usual
noncommutative field theories, is absent in the theories with the
fermionic non-anticommutativity, treated as we suggested. In the second formulation, the Moyal product is polynomial just as in the four-dimensional case, its impact consisting in the appearance of extra vertices, with the UV/IR mixing again absent. Our next step consists in a
more detailed study of quantum corrections. Another interesting
problem is the study of unitarity and causality properties in these new
field theories.

\vspace{1cm}

{\bf Acknowledgements.} This work was partially supported by Funda\c{c}\~{a}o de Amparo \`{a} Pesquisa do Estado de S\~{a}o Paulo (FAPESP) and Conselho Nacional de Desenvolvimento Cient\'{\i}fico e Tecnol\'{o}gico (CNPq). The work of A. F. F. was supported by FAPESP,  project 04/13314-4. A. Yu. P. has been supported by CNPq-FAPESQ, DCR program (CNPq project 350400/2005-9).


\begin{thebibliography}{50}
\bibitem{foot2} A prescription for constructing the supersymmetry generators
compatible with $N=1$ supersymmetry in four dimensional superspace was mentioned in \cite{Seiberg}. In that case, the $N=\frac{1}{2}$ supersymmetry seems to be more advantageous because of the simpler structure of the  generators.
\bibitem{foot} Notice that the supercovariant derivative $D^2_{\a}$ anticommutes with the following extension of the generator $Q^2_{\b}$: $\tilde{Q}^2_{\b}\,=\,Q^2_{\b}-\Sigma^{\mu \rho} \pa^2_{\rho}\pa_{\mu \b}$, with $\{\tilde{Q}^2_{\a},\tilde{Q}^2_{\b}\}=2P_{\a\b}+\Sigma^{\mu\nu}P_{\mu\a}P_{\nu\b}$.
\end{thebibliography}
\end{document}